\documentclass[final,1p,times]{elsarticle} 
\usepackage{graphicx} 
\usepackage{amssymb} 
\usepackage{amsthm} 
\usepackage{amsmath}

\newcommand{\pt}{$p_T$ }
\newcommand{\highpt}{high-\pt}
\newcommand{\raa}{$R_{AA}$ }

\newcommand{\iaacomma}{$I_{AA}$}

\newcommand{\vtwocomma}{$v_2$}
\newcommand{\infinity}{\infty}
\newcommand{\be}{\begin{equation}}
\newcommand{\ee}{\end{equation}}
\newcommand{\eq}[1]{Eq.~(\ref{#1})}

\newcommand{\fig}[1]{Fig.~\ref{#1}}

\journal{Nuclear Physics A} 
\begin{document} 

\begin{frontmatter} 


\title{Shock Treatment: Heavy Quark\\Energy Loss in a Novel AdS/CFT Geometry}

\author{W.\ A.\ Horowitz}

\address{Department of Physics, The Ohio State University\\
191 West Woodruff Avenue, Columbus OH 43210, USA}

\begin{abstract} 
We first note the failures of traditional pQCD techniques as applied to \highpt heavy ion physics and the suggestion of examining the double ratio of charm to bottom nuclear modification factors to generically distinguish between these weak coupling ideas and the strong coupling ideas of AdS/CFT.  In order to gain confidence in the use of AdS/CFT (and to increase the likelihood of falsifying it and/or pQCD) we extend its application to heavy quark energy loss in both thermal and nonthermal media by calculating the string drag in a shock metric.
\end{abstract} 

\end{frontmatter} 




\section{Introduction}\label{sec:introduction}
Despite the early successes of perturbative QCD (pQCD) in describing the highly averaged \highpt light hadron \raa physics at RHIC (see, e.g., \cite{Horowitz:2007su} and references therein), closer examination of a number of experimental observables (e.g.\ \highpt \vtwocomma, nonphotonic electron \raa and \vtwocomma, see, e.g., \cite{Horowitz:2007su}, and even \iaacomma \cite{Nagle:2009wr}) shows that these pQCD techniques do not currently provide a quantitatively consistent picture of all the known data.  Worse, pQCD does not simultaneously describe any two of these quantitatively.  On the other hand the large coupling techniques of AdS/CFT are known to give a reasonable qualitative understanding of a number of RHIC phenomena (see, e.g., \cite{Horowitz:2007su}).  Recent work \cite{Horowitz:2007su} suggests examining the double ratio of charm to bottom $R_{AA}$ as a means of falsifying either the usual pQCD approach or the AdS/CFT one (or both) to \highpt jet quenching.

Specifically the application of AdS/CFT to heavy quark energy loss has shown particular promise \cite{Mikhailov:2003er,Gubser:2006bz}.  Previous work considered a string hanging in the fifth dimension as the representation of a heavy quark in the 4D theory in an empty space metric \cite{Mikhailov:2003er} and in a black hole metric \cite{Gubser:2006bz}, the former to calculate the vacuum energy loss of an accelerating heavy quark and the latter to calculate the energy loss of a heavy quark in a thermalized medium of $\mathcal{N}=4$ SYM plasma.  In this work \cite{Horowitz:2009pw} we take the metric to be that of a shock wave.  We are motivated to do so because we wish to examine the universality of the heavy quark energy loss, hoping to find behavior that will persist in a string dual of QCD.  It turns out that our drag result depends on the typical transverse momentum of the shock medium particles.  When that scale is related to a temperature, we exactly reproduce the previous AdS/Schwarzschild results; however the shock can describe matter with any isotropic distribution of momentum, and we therefore conclude that we have generalized the string drag calculation.

\section{Shock Metric, Drag Force, and Limits of the Calculation}\label{sec:metric}
We will work in an asymptotic AdS$_5$ space with metric
 \cite{Janik:2005zt}
\begin{subequations}
\begin{align}
\label{lcmetric}
ds^2 & \equiv \, G_{\mu\nu} \, d x^\mu \, d x^\nu \, = \,
\frac{L^2}{z^2} \, \left[ -2dx^+dx^- + 2 \, \mu \, z^4 \, \theta(x^-)
  \, dx^{-2} + dx_\perp^2+dz^2 \right] \\
\label{metric}
& = \frac{L^2}{z^2} \, \left[ -\left(1- \mu \, z^4 \, \theta(x^-)
  \right) \, dt^2 - 2 \, \mu \, z^4 \, \theta(x^-) \, dt \, dx +
  \left(1+\mu \, z^4 \, \theta(x^-)\right)dx^{2}+dx_\perp^2+dz^2
\right],
\end{align}
\end{subequations}
where we have used the $x^\pm=(t \pm x)/\sqrt{2}$ normalization of
light-cone coordinates with $x = x^3$ and dropped the $d\Omega_5^2$
standard metric of the five-sphere in AdS$_5\times S^5$. As usual $L$
is the radius of the $S^5$ space, and $dx_\perp^2 = (d x^1)^2 + (d
x^2)^2$ is the transverse part of the metric.


The dual of a finite mass quark in the fundamental representation in AdS/CFT is an open Nambu-Goto string terminating on a D7 brane at $z=z_M=\sqrt{\lambda}/(2\pi M_q)$ \cite{Gubser:2006bz}; the other end of the string ends on the stack of $N_c$ color branes at $z=\infinity$.  The test string action is
\be
\label{ngaction}
\begin{split}
  S_{NG} & = -T_0\int d\tau d\sigma\sqrt{- g}, \\
  g \, & = \, \mbox{det} \, g_{ab}, \ \ \ g_{ab} = G_{\mu\nu} \,
  \partial_a X^\mu \, \partial_b X^\nu,
\end{split}
\ee
where $G_{\mu\nu}$ is the spacetime metric of \eq{metric}, Greek
indices refer to spacetime coordinates, and Latin indices to
worldsheet coordinates.  $X^\mu=X^\mu(\sigma)$ specifies the mapping
from the string worldsheet coordinates $\sigma^a$ to spacetime
coordinates $x^\mu$.  The backreaction of the fundamental string
($\mathcal{O}(N_c)$) is neglected as compared to the
$\mathcal{O}(N_c^2)$ contributions from the adjoint fields of
$\mathcal{N}=4$ SYM \cite{Gubser:2006bz}.

\begin{figure}[!htb]
\centering
\includegraphics[width=.6\textwidth]{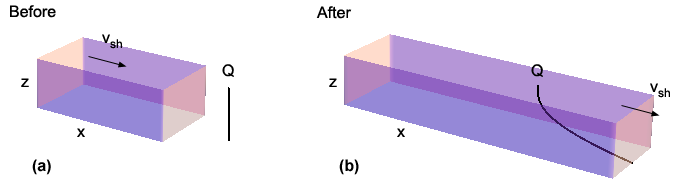}
\caption{\label{setup}
(Color online) An illustration of the shock of the metric \eq{lcmetric} colliding with a heavy quark $Q$ in its rest frame. (a) before the collision the Nambu-Goto string, \eq{ngaction}, hangs straight down; (b) afterwards the string is dragged by the shock behind its endpoint.
}
\end{figure}

\begin{figure}[!htb]
\centering
\includegraphics[width=.6\textwidth]{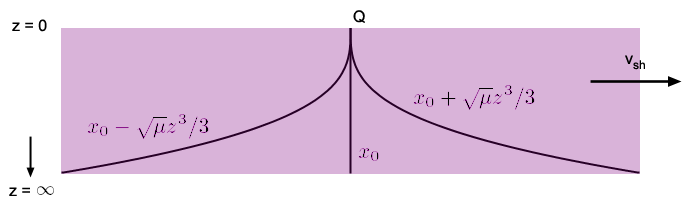}
\caption{
The three solutions to the static equations of motion, $x(t,z)=\xi(z)=x_0,\, x_0\pm\sqrt{\mu}z^3/3.$}
\label{solns}
\end{figure}

Varying the action, \eq{ngaction}, yields the equations of motion: $\nabla_a P^a_\mu = 0,$ $P^a_\mu = \pi^a_\mu/\sqrt{-g} = -T_0 \, G_{\mu\nu} \, 
\partial^a X^\nu$,
where $\nabla_a$ is the covariant derivative with respect to the
induced metric, $g_{ab}$, and the $\pi^a_\mu$ are the canonical
momenta, $\pi^a_\mu = -T_0 \, \partial\sqrt{-g}/\partial (\partial_a X^\mu)$.
%
Limiting our attention to only the nontrivial directions of $t$, $z$, and $x$, $X^\mu(\sigma)$
maps into the $(t,x,z)$ coordinates; choosing the static gauge,
$\sigma^a=(t,z)$, the string embedding is described by a single
function, $x(t,z)$.  We find the equations of motion for $x(t,z)$ by plugging it into the action and then varying with respect to $t$ and $z$.  Denoting $\partial_t x = \dot{x}$ and $\partial_z x = x'$ we find $-g = L^4\big(1+x'^2-\dot{x}^2-\mu z^4 (1-\dot{x})^2\big)/z^4$ and $\partial_t\big((\mu z^4 - (1 + \mu z^4)\dot{x})/z^4\sqrt{-g}\big)+\partial_z\big(x'/z^4\sqrt{-g}\big)=0$.
%
If we assume a static solution ansatz $x(t,z) = \xi(z)$ in a shock that fills spacetime the equations of motion become $\partial_z (\xi'/z^4 \sqrt{-g})=0$.
Setting the quantity in the parentheses equal to an integration constant $C$ one may solve for $\xi'$: $\xi'=\pm C z^2 \sqrt{(1-\mu z^4)/(1-C^2 z^4)}$.
There are two cases of interest for a string hanging from small $z$ to $z=\infinity$: $C=0$ and $C\ne0$.  For the latter case $C$ is set by requiring reality of the solution.  At small $z$ both the numerator and denominator of the equation for $\xi'$ are positive; at large $z$ both are negative.  For their ratio to always be real $C = \sqrt{\mu}$.  Trivial integration yields $\xi(z) = x_0\pm\sqrt{\mu}z^3/3$.
It is interesting to note that the
near-boundary expansion of the static quark solution in the black
hole metric (with horizon at $z=z_h$) is $x(t,z)\approx x_0+vt\pm
z^3/(3z_h^2)$.  For $C=0$ the solution $\xi = x_0$ is immediately found; the string hangs straight down.  We plot the three solutions in \fig{solns}.  

One resolves the sign ambiguity in the pair of solutions for $C=\sqrt{\mu}$, resulting from the time-reversal symmetry of the problem, by taking the physical one, which trails behind the quark and has the positive sign; the negative sign solution has the string ``trailing'' in front of the quark.  
Additionally plugging the straight solution back into the action \eq{ngaction} gives $S = -T_0 \int dt \int_{z_M}^\infinity dz \sqrt{1-\mu z^4}/z^2$.
The IR region of the $z$ integration contributes an infinite imaginary part to the action; we interpret this solution as infinitely unstable that would immediately decay into the physical, trailing string solution.

The drag force on the heavy quark in the SYM theory corresponds to the
momentum flow from the direction of heavy quark propagation down the
string, $dp/dt = -\pi^1_x$.  From the canonical momenta, in the rest frame of the heavy quark, $dp/dt = - \pi^1_x = \sqrt{\lambda \mu}/2\pi$.
Formally the metric, \eq{lcmetric}, has the shock on the light cone; thinking of this as an approximation to a shock nearly on the light cone one has a well-defined rest frame for the shock.  This then is the lab frame, where we want to know to momentum loss of the heavy quark, and also allows us to relate $\mu$ to properties of the medium.  

\begin{figure}[!htb]
\centering
\includegraphics[width=.6\columnwidth]{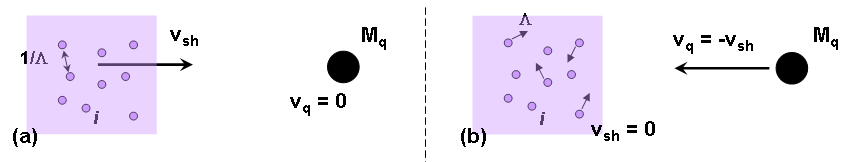}
\caption{\label{frames}
(a) Shock and quark system as viewed in the heavy quark rest frame. (b) In the shock rest frame (the lab frame).
}
\end{figure}

Following \cite{Albacete:2008ze}, we assume the medium is made up 
of $N_c^2$ valence gluons of the $\mathcal{N}=4$ SYM fields; 
see \fig{frames}.  If in the rest frame of the medium the particles 
are isotropically distributed with a typical momentum of order $\Lambda$---with 
associated inter-particle spacing of order $1/\Lambda$---then the 00 component of the 
stress-energy tensor in the rest frame of the shock is $\langle T'_{00} \rangle \, \propto \, N_c^2 \Lambda^4$,
where primes denote quantities in the rest frame of the medium and
proportionality is up to a constant numerical factor.  Transforming into lightcone
coordinates, boosting into the rest frame of the heavy quark, and dropping the $\mathcal{O}(1)$ constant of proportionality, yields $\langle T_{--} \rangle = N_c^2 \Lambda^4 \gamma^2 = N_c^2 \Lambda^4 (p'/M)^2$,
where we assumed ultrarelativistic motion for the heavy quark in the
medium rest frame, $p' \simeq M\gamma$.  Comparing this with the energy momentum tensor found above we read off $\mu = \pi^2 \Lambda^4 (p'/M)^2$.

To rewrite $dp/dt$ in terms of the momentum
and time in the medium rest frame, $dp'/dt'$, note that $dp/dt$ is the 3-vector
component of the force 4-vector in the quark rest frame: $f^x \equiv dp/d\tau = dp/dt$. $\pi^1_t = 0$, and hence $f^t = 0$, 
and the 4-force boosted into the shock rest frame is
$f'^x = -\gamma f^x = -\gamma dp/dt$, where the negative sign comes from boosting into a frame moving in the opposite direction; see \fig{frames}.  From the definition of the 
4-force we also know that in this frame $f'^x \equiv dp'/d\tau = \gamma dp'/dt'$. Hence we find that $dp/dt = -dp'/dt'$. Using these we find our main result, 
\begin{align}
  \label{momgaintwo}
  \frac{d p'}{d t'} \, = \, - \frac{\sqrt{\lambda}}{2} \,
  \frac{\Lambda^2}{M_q} \, p'.
\end{align}
If we take the typical medium particle momentum $\Lambda = \sqrt{\pi}T$ then our result exactly reproduces that of the black hole metric, $d p' / d t'
= - \pi \sqrt{\lambda} \, T^2 \, p'/ (2 \, M_q)$
\cite{Gubser:2006bz}. 

One may readily derive a ``speed limit'' of applicability of this formalism in the case of a static heavy quark solution.  The momentum lost to the medium must be balanced by a momentum input from a Born-Infeld derived electric field on the D7 brane, which has a natural cutoff at the energy scale of heavy quark pair production \cite{Gubser:2006bz}.  For the case of a heavy quark allowed to slow down of its own accord one may examine the speed limit of a point particle traveling at $z_M$ \cite{Gubser:2006nz}.  This gives the same limit as the one from the BI action, but it is not entirely clear that it implies a restriction on the string whose action remains real.  Nevertheless one may derive it in the shock metric; while the shock does not support a black hole horizon the local speed of light has a nontrivial $z$ dependence $(\mu z^4-1)/(\mu z^4 + 1) \le v \le 1$.
Setting $v=0$ for our static quark gives $\mu \, z_M^4 \le 1$
for the limit. Using $z_M = \sqrt{\lambda}/(2\pi M_q)$ and $\mu=\pi^2\Lambda^4\gamma^2$, along with $\Lambda = \sqrt{\pi} \, T$, we obtain $\gamma \le (4 \, \pi \, M_q^2)/(\lambda \, \Lambda^2) = (4 \, M_q^2)/(\lambda \, T^2).$
We note that this limit is identical to
that for the BH metric \cite{Gubser:2006bz,Gubser:2006nz}. 

\section{Conclusions}\label{sec:conclusions}
Advances in experimental heavy ion physics holds out the promise of testing both the usual pQCD and the novel AdS/CFT formalisms.   
A rigorous understanding of the theoretical error and regimes of applicability associated with the predictions of these theories is required in order to falsify one or both experimentally.  In this work we generalized AdS/CFT heavy quark energy loss to media of any isotropic distribution of momentum.  We found that, just as in pQCD, the form of energy loss is independent of the thermal or nonthermal nature of the medium.



\section*{Acknowledgments} 
This work was supported by the Office of Nuclear Physics, of the Office
of Science, of the U.S.\ Department of Energy under Grant No.\ 
DE-FG02-05ER41377.

\end{document}